\documentclass[sigconf]{acmart}
\usepackage{microtype}
\usepackage{graphicx}
\usepackage{subfig}
\usepackage{geometry}
\usepackage{textcomp}
\usepackage{multirow}
\usepackage{framed}
\usepackage{comment}
\usepackage{pgfplots}
\usepackage{amsmath}
\usepackage{physics}
\usepackage{url}

\usepackage{hyperref}

\newcommand{\eat}[1]{}

\title{The Jevons Paradox In Cloud Computing: A Thermodynamics Perspective}
\author{Prateek Sharma}
\affiliation{
  \institution{Indiana University Bloomington}
  \country{USA}}
\email{prateeks@iu.edu}

\begin{document}

\begin{abstract}
How do we explain the simultaneous growth in energy efficiency of cloud computing and its energy consumption?
The Jevons paradox provides one perspective of this phenomenon.
However, it is not clear or obvious \emph{why} the Jevons paradox exists, and \emph{when} is it applicable.
To answer these questions, we seek inspiration from thermodynamics, and model the cloud as a thermodynamic system.
We find that system growth, due to the revenue generation of cloud platforms, is a key driver behind energy consumption.
This thermodynamic model provides energy consumption insights into modern hyperscale clouds, and we validate it using data from Meta and Google.
Our investigation points to the necessity of future work in new and meaningful efficiency metrics, implications for future applications and edge clouds, and the need for studying system-wide energy and sustainability. 
\end{abstract}

\maketitle

\section{Introduction}
\label{sec:intro}

The information and communication industry, and cloud computing in particular, are significant energy consumers on the global scale~\cite{jones_how_2018}. 
It is estimated that cloud platforms consume around 1\% of the global energy, with a significant growth predicted due to AI, IoT, and other emerging cloud-based applications~\cite{siddik_environmental_2021, gelenbe_2015, freitag_real_2021}. 

Cloud platforms and their data centers are incredibly efficient: hyperscale data centers achieve a PUE of close to 1.1, meaning that only 10\% of the energy is consumed by ``non-IT'' components such as cooling etc.
Furthermore, their large size and use of virtualization provides statistical multiplexing and bin-packing opportunities, allowing the servers to be highly utilized compared to traditional non-virtualized bare-metal enterprise clusters. 
The high profit margins and revenue of clouds also allows them to invest in green renewable energy and purchase various forms of carbon offsets~\cite{maji2023untangling}. 
Load management techniques such as carbon-aware spatial and temporal workload shifting (such as running energy-hungry jobs during day-time) will improve the efficiency even further~\cite{radovanovic_carbon-aware_2023,sukprasert_quantifying_2023}.

\begin{figure}
  \centering
  \includegraphics[width=0.4\textwidth]{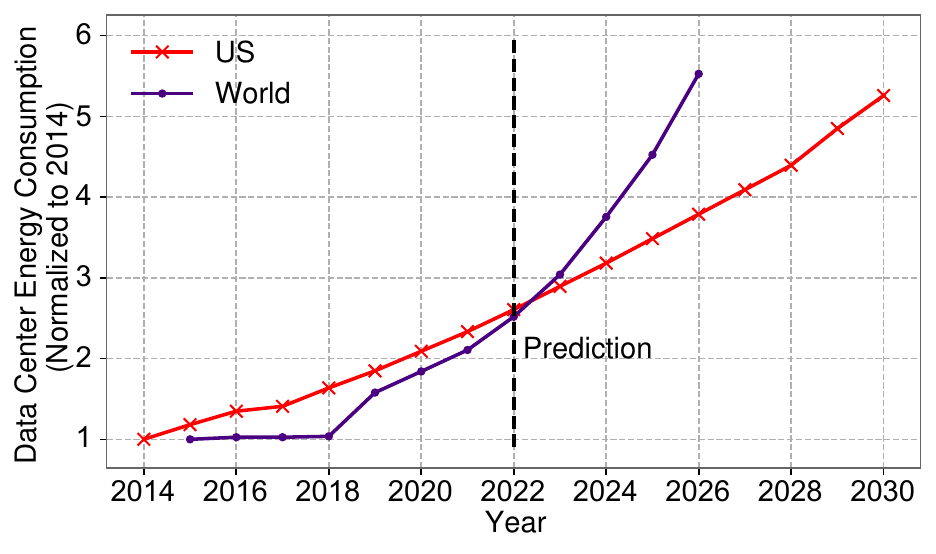}
  \vspace*{-7pt}
  \caption{Energy consumption of data centers in the US and the world is increasing, inspite of all efficiency advances. Data from~\cite{mckinsey_dc,iea24,shehabi2016united}.}
  \label{fig:us-world}
    \vspace*{-12pt}
\end{figure}

All these factors, among others, are frequently touted by cloud providers~\cite{ms_sust_calc} as reasons why cloud hosting is ``green'' and beneficial in reducing the carbon footprint of applications. 
Yet, inspite of these long-standing serious efforts to improve efficiency, the energy consumption of cloud computing continues to rise: for example, Google's data center consumption increased by $3.7\times$ from 2016 to 2022, and world-wide data center energy has increased by $2.5\times$ in the same period (Figure~\ref{fig:us-world}).

The \textbf{Jevons Paradox} offers one explanation of this phenomenon.
In many scenarios, efficiency in using a resource leads in the medium/long term to an increased use of that resource. 
First observed and articulated by William Henry Jevons in 1865 in the context of early steam engine efficiency and coal consumption,  this ``paradox'' is a key maxim in environmental science and economics. 
It is considered a paradox because improving efficiency is obviously a beneficial pursuit, yet the end results are the opposite of the intended goal. 
Taken in its most pessimistic form, it points to the futility of pursuing efficiency and reducing production costs, which accelerates consumption. 
In other cases, it is useful as a ``warning'' that improving efficiency alone is insufficient, and that a broader holistic ``systems-view'' be considered.

\emph{Why does this efficiency vs. consumption paradox occur?}
To answer this question, we use a \textbf{thermodynamics} perspective, which we adapt from prior work in~\cite{garrett_no_2012}.
We seek a phenomonological model, with intuitive mechanistic explanation, and which is simple enough to explain highly complex cloud systems. 
Thermodynamics concerns itself with the energy consumption, efficiency, and evolution of large systems, and statistical mechanics and thermodynamics can provide ``coarse-grained'' models for explaining complex phenomenon.

From a thermodynamics perspective, the Jevons paradox can be explained by a key feedback loop: \emph{as a system improves in efficiency, it expands in size, consuming more energy.}
Such models have been applied to explain the energy trends of large systems such as the global economy, biological systems, etc~\cite{garrett2011there,garrett_lotkas_2022}. 
Thus it is a natural question to see if we can examine cloud computing through the thermodynamic lens, and understand the deeper relationship between efficiency and growth of modern digital infrastructure. 

One of our main contributions is to apply and explain these thermodynamic models in the context of cloud computing. 
Our models are simple and rest on two key concepts: the feedback loop between efficiency and size, and the role of network effects and ``system inertia''. 
We consider the entire cloud ecosystem (including users of cloud-based applications), and are able to model the energy consumption, efficiency, user-base, and financial metrics such as the revenue and capital expenditure. 
By stepping back from the minitua of cloud scheduling and operations, the coarser-grained thermodynamics model allows us to relate the above metrics in new and intuitive ways, and provide a new way of understanding energy consumption.

One of the main challenges is developing the right model abstractions so that cloud metrics can be combined with financial and thermodynamic ones, while ensuring these metrics are easily measurable and publicly available (Section~\ref{sec:model}). 
We \textbf{validate} our models using data from two cloud systems: Meta and Google. 
Using publicly available energy and financial data, we find that our thermodynamics approach is accurately able to model the increase in energy consumption (Section~\ref{sec:eval}).
The Jevons Paradox continues to hold for the hyperscale cloud platforms, and that energy expenditure and system size both continue to grow in accordance with the thermodynamics model. 
Finally, in Section~\ref{sec:discuss}, we highlight how our modeling approach shows the importance of efficiency metrics for the broader community, the role of decentralization and edge clouds, and argue for new a systems and energy-centered approach to sustainability for cloud computing.

\section{Background}
\label{sec:bg}

\subsection{Cloud Sustainability}

Cloud computing is some of the most energy dense activities and also the fastest growing energy consumer~\cite{iea24}.
The large electricity costs provide a natural incentive for cloud providers to improve the data center efficiency, and as a result, the power usage efficiency (total power divided by IT power) of hyperscale cloud platforms is around 1.1. 
To reduce carbon emissions, resource management techniques such as temporal and spatial workload shifting~\cite{sukprasert_quantifying_2023} and data center demand-response~\cite{lin_adapting_2023, radovanovic_carbon-aware_2023} can be used.
By using renewable energy sources and purchasing carbon offsets, the operational carbon footprint of cloud platforms is low~\cite{bashir2023sustainable, maji2023untangling}. 
Recently, embodied emissions (also known as Scope 3 emissions) due to the manufacturing of server equipment has emerged as an important metric and optimization criteria~\cite{gupta_act_2022, hanafy2023war}. 

While carbon optimizations are important, we take an energy-centered view of the environmental footprint of the cloud computing ecosystem.
Carbon is only one part of the ``waste'' of the cloud system.
Energy consumption is a fundamental and universal metric which directly affects resource depletion of water, minerals, and is the main drivers of environmental pollution.
In contrast, carbon emissions are difficult to measure consistently, and the different accounting practices and the uncertain benefits of offsets~\cite{xu_system-level_2024} make them a less reliable metric for system-wide analysis.

\subsection{Jevons Paradox}

The paradox was first put forth in the 1865 book, The Coal Question: 
``It is wholly a confusion of ideas to suppose that the economical use of fuel is equivalent to a diminished consumption. The very contrary is the truth.''
It has a rich history in economics and ecology~\cite{alcott_book}, where it is often referred to as ``the rebound effect''.
One of the main driving mechanisms is that energy efficiency improves cost efficiency, which drives up consumption.
This was also originally observed by Jevons later in the same book:
``As a rule, new modes of economy will lead to an increase of consumption''. 

Despite its central role in the governing dynamics of energy consumption and economic growth, it has seen relatively little formal analysis.
For global energy consumption and economic growth, a thermodynamic model was developed by Garrett in~\cite{garrett2011there, garrett2012modes, garrett_lotkas_2022, garrett_long-run_2015}.
Our model is largely based on the above thermodynamic models which we adapt and apply to cloud environments.
More broadly, the works of Georgescu-Roegen~\cite{georgescu1975energy, georgescu1971entropy} in bio-economics have also used thermodynamic principles such as entropy to understand and model economic growth.

\section{Cloud Thermodynamics Model}
\label{sec:model}

\begin{figure}
  \centering
  \includegraphics[width=0.4\textwidth]{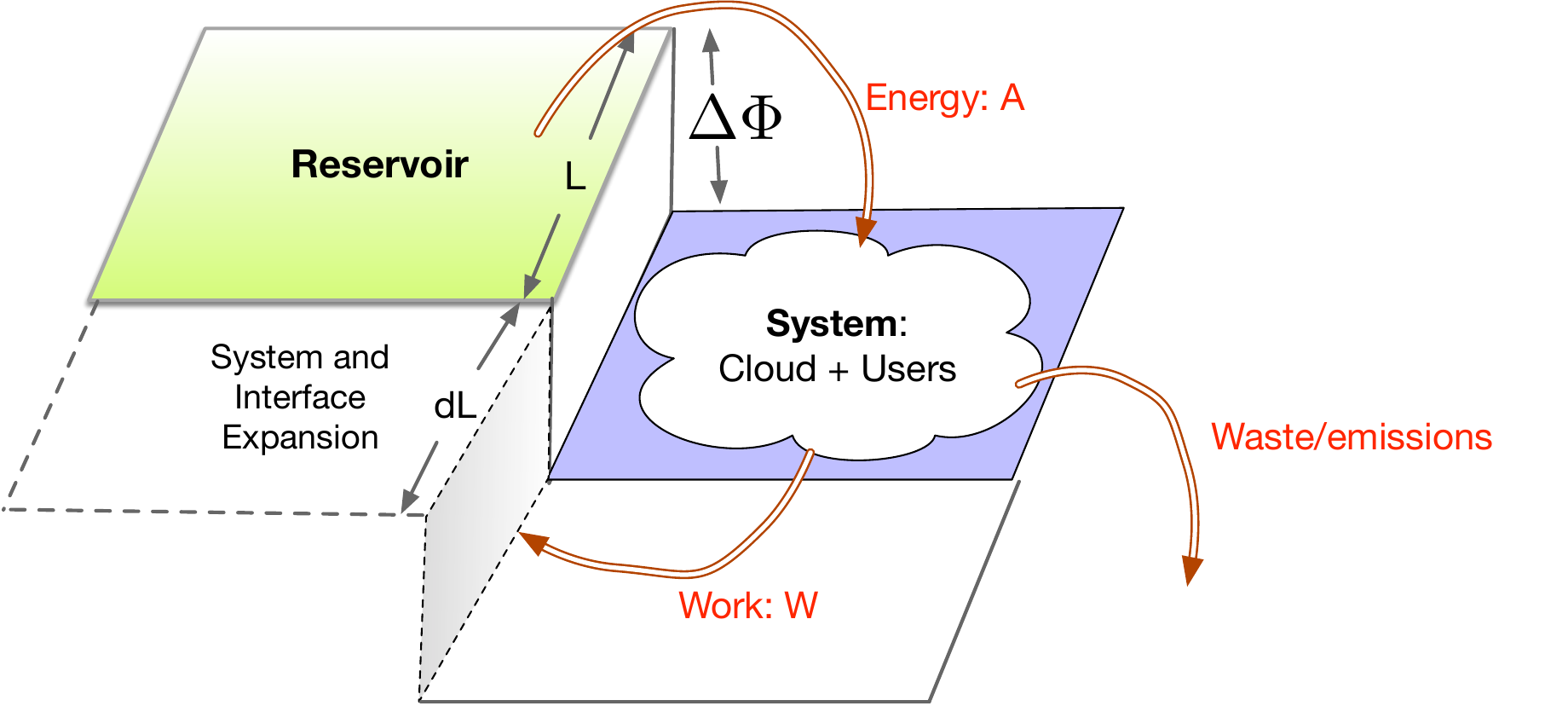}
  \caption{Thermodynamic model of cloud computing. Energy and matter flow from the higher-potential reservoir. Work (revenue) expands the system's interface and increases the energy consumption further.}
  \label{fig:sys-model}
\end{figure}

We propose to use a coarse-grained energy model of cloud platforms, which is similar to thermodynamics models of similar complex systems such as global economy and energy use~\cite{garrett2011there, georgescu1971entropy}. 
We view the \emph{cloud data centers and users} as a large ensemble and a complex system. 
It is an \emph{open} system with porous boundary with the environment, through which energy and matter can flow. 
The system has a boundary with the reservoir which is the source of the energy and material flux (Figure~\ref{fig:sys-model}). 

Conventional thermodynamics models are composed of ensembles of a very large number of discrete particles with extrinsic aggregate properties such as entropy, temperature, pressure, etc. 
The cloud system is more abstract and is the collective aggregate of all the servers, clients, and communication networks that comprise the cloud  and users ecosystem. 
Since hyperscale cloud platforms such as Google and Meta comprises of millions of computing devices and billions of users, we can treat it as a large-enough ensemble for thermodynamic modeling.

The system (Figure~\ref{fig:sys-model}) is able to draw energy and matter from the reservoir because of a potential difference $\Delta \Phi$, which captures the energy and material potentials. In classical thermodynamics, this is the Gibbs potential $\Delta G$, which also captures the chemical potential (difference in material density). 
Based on this potential, the system consumes energy at rate $A(t)$ (we will omit time parameters for clarity of exposition). 
The system's ability to extract energy from the reservoir also depends on the ``conductivity'' $\alpha$, and the size of the interface boundary, $L$.
This interface boundary is one of the key parameters which influences how such open systems interact with the environment, and crucially, can also change over time, as we shall model.

\noindent \textbf{Energy consumption} is governed by three factors:
\begin{equation}
  \label{eq:A}
  A = \alpha \cdot L \cdot \Delta \Phi, 
\end{equation}
where $A$ is the electricity consumption of the cloud data centers and the client-side consumption.  
The latter comprises of the energy consumed in using the cloud-based applications and services, as well as the energy to communicate over the Internet.
All quantities are a function of time, i.e., $A(t), L(t), \Delta \Phi(t)$, and we omit the explicit time parameter for brevity.
For the cloud system, these quantities are often measurable or available in coarse granularities---for example, sustainability reports from hyperscale cloud provide yearly aggregates of energy consumption, in which case $A(t)$ represents the yearly energy consumption.  

\noindent \textbf{Work} is defined as raising the system-wide potential of \emph{some} interest. 
The work done per unit of energy consumed determines the conventional notion of system efficiency ($\varepsilon$), and is:
\begin{equation}
  \label{eq:W}
  W = \varepsilon \cdot A  
\end{equation}

\emph{Work and efficiency are challenging to define in cloud contexts, and depend on the chosen system boundaries.}
In conventional data-center level metrics such as PUE are popular, however, they do not capture the utility of the work performed, and how the work influences the rest of the system and the environment.
Consider a perfectly efficient data center (with PUE equal to 1.0) that runs idle servers or performs computation whose output is never communicated or used.
From a cloud system perspective, we would consider this system's ``output'' as zero. 

Thus, metrics for work have to reflect the \emph{aggregate} raising of some system-wide potential, and the output must be valuable to the rest of the system.
Based on these requirements, we claim that \emph{cloud revenue} is a suitable work metric.
Cloud platforms are thus systems for converting energy into revenue, and the efficiency metric $\varepsilon$ is thus in units of dollars/Joule.
In terms of thermodynamic modeling, using revenue as the work metric has many advantages: it is well-understood and publicly available, and is the reason-detrie of cloud platforms, so is also ``incentive compatible'' from an economics perspective. 
We discuss additional and finer-grained work and efficiency metrics later in Section~\ref{sec:discuss}. 

In isolation, improving the efficiency of the system $\varepsilon$ will result in lowered total energy consumption for the same amount of work.
However, this is often not the case, because as the system becomes more efficient, it can more easily \emph{grow}, and increase its interface with the environment, and thus consume more energy. This is the key idea driving the paradox.

\noindent \textbf{Work-induced system growth.}
Energy consumption and work are the drivers behind Jevons paradox.
Since work raises the potential of the system, it leads to an increase in the size of the interface $L$ with the environment (i.e., the reservoir).
In cloud context, revenue is often re-invested in increasing the system size, by building more data centers, software and algorithms for increasing user engagement, etc.
Both the above examples represent growth of the system frontier.
Work thus captures how a system grows: 
\begin{equation}
  \label{eq:W2}
    W = \dv{L}{t} \cdot \Delta \Phi. 
\end{equation}

$L$ represents the size of the boundary between system and the environment. 
For three dimensional physical systems, it is proportional to $L = N_S^{1/3}*N_R$, where $N_S$ is the system size and $N_R$ is the reservoir size.

This leads us to the first intuitive explanation for Jevons paradox: as the system efficiency increases and more work gets done, the system increases in size and the frontier length $L$ increases (Equation~\ref{eq:W2}), but this also increases energy consumption, since $A$ is proportional to $L$ as per Equation~\ref{eq:A}. 
We present a deeper look at the dynamics of work and energy next.

\noindent \textbf{Dynamics of growth.}
To understand how the system's energy use can increase with time, consider the following set of equations which are derived from the base definitions: 
\begin{equation}
  \label{eq:L}
  \varepsilon = \frac{W}{A} = \frac{1}{\alpha L} \dv{L}{t} = \frac{1}{\alpha} \dv{\ln L}{t}
\end{equation}

Thus, the efficiency is proportional to the rate of growth of the natural logarithm of the frontier length. 
We define $\eta = \alpha \cdot \varepsilon$, and it represents the rate of growth of the system.
Simplifying the above equations, we get:
\begin{equation}
  \label{eq:L2}
  \dv{\ln L}{t} = \alpha \varepsilon = \eta 
\end{equation}

A constant $\eta$ will lead to an exponential growth in size and energy.
This depends on the reservoir's ability to provide energy and material at constant rates (i.e., $\Delta \Phi$ remains constant).
If we consider the depletion of the reservoir and a decrease in the potential, then we can obtain sigmoid or S-shaped rate of growth.
The growth rate is related to the entropy ($S$) of the system: $\eta =  \frac{d ln S}{dt}$, for the perfect constant-potential reservoir~\cite{garrett2012modes}.

\noindent \textbf{Energy growth.}
We can now combine the previous equations~\ref{eq:A},~\ref{eq:W},~\ref{eq:W2}, to get the rate of energy growth: 
\begin{equation}
  \label{eq:A2}
  \dv{A}{t} = \alpha \cdot \Delta \Phi \dv{L}{t} = \eta A 
\end{equation}
This will lead to the exponential energy growth, and an increase in efficiency ($\varepsilon$) only increases the rate of growth:
\begin{equation}
  \label{eq:A-int}
  A(t) = e^{\eta t}.
\end{equation}

\noindent \textbf{Embodied emissions} have become an increasingly important sustainability metric, which can also be estimated from our model. By definition, the emissions embodied in computing hardware are a result of the cumulative energy and material flows. If the system is expanded to include hardware manufacturing and distribution entities, then the total embodied emissions is proportional to the historical cumulative energy usage. That is, $\text{Embodied(t)} = \int_0^t A(x) dx$. This quantity is proportional to the length $L$ of the system interface---providing another interpretation of embodied emissions: they represent the size of the interface between the cloud system and the environment.

\noindent \textbf{System inertia and memory.}
As per the previous equation, the rate of energy increase is  proportional to the present energy usage.
The present energy consumption is thus proportional to the cumulative and historical energy consumption $A(x) = \int_0^x \eta A(x) dx$.
This means that our system has inertia and memory: the current consumption is based on historical energy and material flows into the system.

The role and importance of inertia is also observed in the growth of global wealth and energy consumption~\cite{garrett2011there}.
It has been empirically shown that the global energy consumption $A$ is proportional to the cumulative civilizational wealth.
In our cloud system model, the cumulative revenue is: $C(t) = \int_0^{t} W(x) dx$. 
Using Equation~\ref{eq:W2}, we get: $A = \alpha C$, which indicates that the current energy consumption is proportional to the cumulative revenue of the cloud platform. 

System inertia may lead to a better understanding of the effectiveness of efficiency improvements. 
As systems grow and expand, they accumulate inertia.
The ``networks'' comprising the system become stronger and larger, resulting in more energy consumption. 
As individual nodes of the system become more efficient, they permit and require bigger energy and material flows (unless constrained by internal or external conditions).
Thus, we see the effect where data center efficiency improvements leads to more revenue, resulting in new energy-hungry AI applications, etc.
The Jevons paradox requires large but still growable networks. Cloud systems fulfill both these requirements, and thus we see the exponential growth in energy usage inspite of all the ``point'' efficiency improvements inside data centers.

\section{Model Evaluation}
\label{sec:eval}

\emph{Can the thermodynamic model explain cloud growth?}

\noindent \textbf{Methodology.}
We use public data from two hyperscale computing providers, Meta and Google. 
For the energy data, we use the electricity consumption provided in the sustainability reports over a period of 2016--2022~\cite{fb-2023-report, goog-2023-report}. 
For revenue and capital investments, we use financial data from the public SEC filings. 
We only consider the cloud provider as part of the system, and not the users, since it is hard to get reliable estimates of how much energy client devices consume when accessing the cloud services.

We are interested in evaluating how closely do the model predictions fit the real data, considering that the simple thermodynamic model is being studied for the first time in the cloud context.

\begin{figure}
  \centering
  \includegraphics[width=0.45\textwidth]{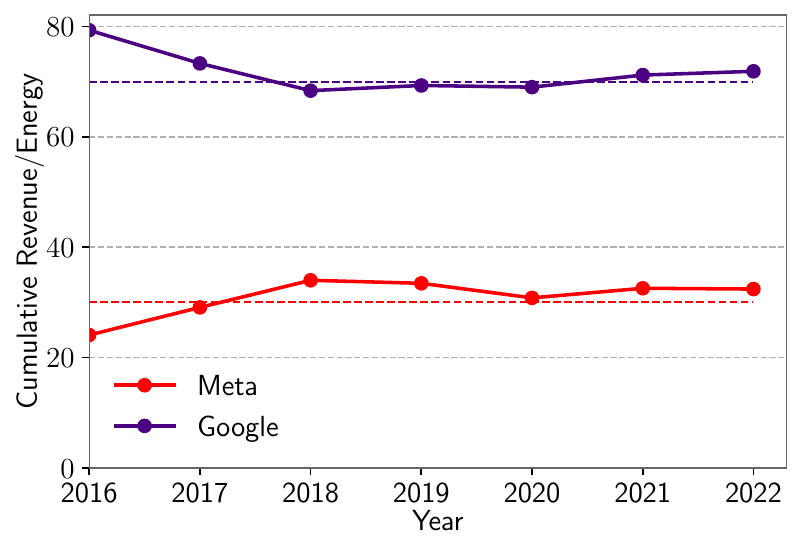}
  \vspace*{-7pt}
  \caption{Data from Meta and Google shows that energy use is proportional to the cumulative revenue, which is the central prediction of the thermodynamics model (dashed lines).}
  \label{fig:CRE}
    \vspace*{-7pt}
\end{figure}

\noindent \textbf{Energy and Cumulative Revenue.}
As derived in the previous section, the energy consumption is proportional to the cumulative work (i.e., $A = \alpha C$). 
This is shown for Meta and Google in Figure~\ref{fig:CRE}, and we see that the model prediction holds true in both the cases.
We have used the historical revenue of both these companies (going past 2016), since it is crucial in determining the system inertia.     

This result implies that both these cloud systems are operating in the exponential growth phase and able to tap into constant-potential reservoirs of energy and material. 
Thus the energy consumption will only continue to rise with revenue.
The longer cloud platforms operate, the more inertia they will accumulate, which will lead to a risk of getting ``trapped''.

\noindent \textbf{System Expansion.}
The other key aspect of the thermodynamics model is the mechanism of system interface expansion which drives the energy usage.
We propose two metrics for capturing and defining the interface size.

\begin{figure}
  \centering \includegraphics[width=0.4\textwidth]{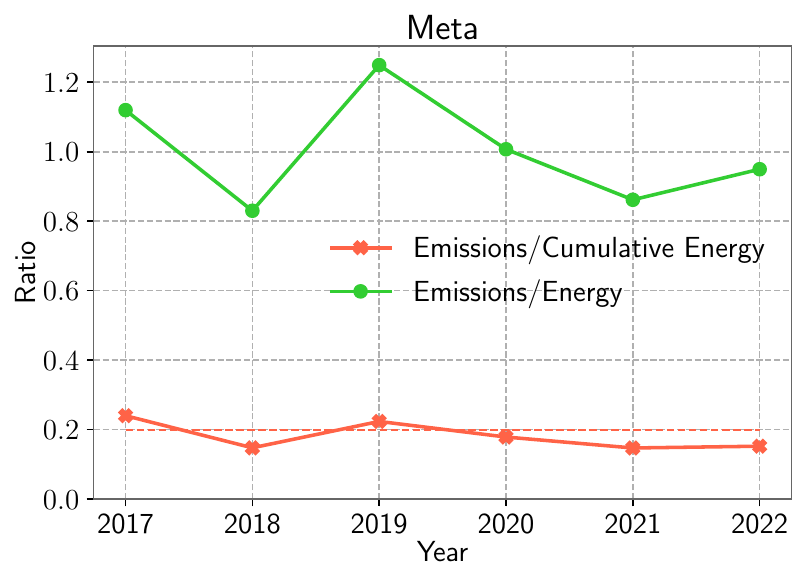}
  \vspace*{-7pt}
  \caption{Embodied emissions can be a good proxy for system size, and have a constant relation with the cumulative energy use, as predicted by the model.}
  \vspace*{-7pt}
  \label{fig:fb-emissions}
\end{figure}

The first proxy for interface size is the cloud \emph{carbon emissions}, including the operational and embodied carbon.
In particular, the embodied emissions captures the material flux into the system (e.g., servers).
Thus, we use the Scope 3 emissions as one of the interface size metrics.
Based on our model, we should expect that interface size (i.e., emissions) are proportional to the cumulative energy usage.
That is, $L(t) \propto \int_0^t A(x) dx$, as per Equation~\ref{eq:W2}.
For Meta, this is shown in Figure~\ref{fig:fb-emissions}, which confirms the proportional relation between the emissions and cumulative energy.
The figure also shows the importance of considering the cumulative and historical energy, since the relation between energy and emissions is \emph{not} stable and not amenable to modeling.

A similar analysis for Google data centers fails to show the proportional relation between Scope 3 emissions and cumulative energy, indicating that emissions alone may not be a universal size proxy.
We observe a steady, unexplained decrease in the claimed Scope 3 emissions, even as new data centers are being built.
Because embodied emissions accounting is non-standard and can change over time, using it as a size-proxy is challenging in general. 

\begin{figure}
  \centering
  \includegraphics[width=0.4\textwidth]{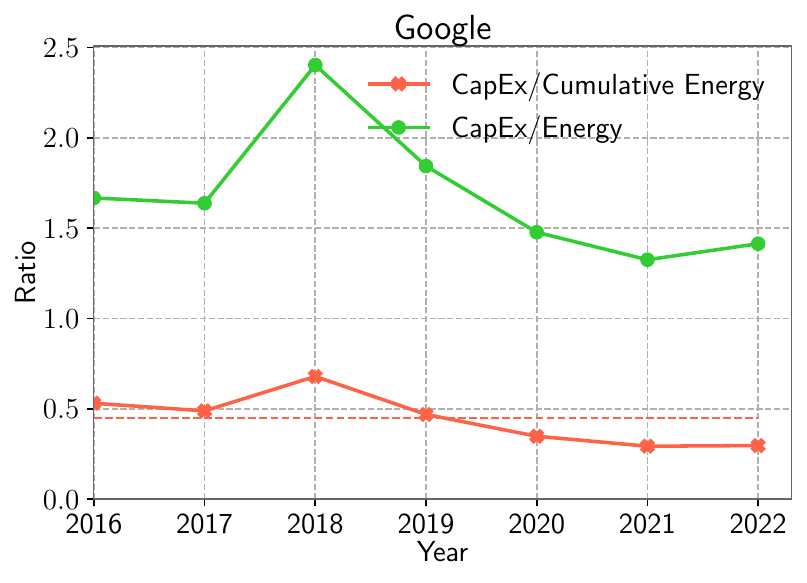}
  \vspace*{-7pt}
  \caption{The Capital Expenditure (CapEx) used for building data centers can also be used as a proxy for system size.}
  \label{fig:google-capex}
  \vspace*{-7pt}
\end{figure}

The second size proxy we can use is the capital expenditure, which primarily goes into the data centers.
Thus, we should expect the capital expenditure to be proportional to the cumulative energy consumption.
This is shown in Figure~\ref{fig:google-capex}, which shows the proportionality.
This further confirms the model predictions and validates the use of capex as a proxy for system interface size. 

\noindent \textbf{Model-based back-casting.}
For the analysis presented in Figure~\ref{fig:google-capex}, historical energy use was difficult to obtain with public data sources, since the sustainability reports only go as far back as 2016.
To estimate the cumulative historical (pre-2016) energy usage, we use our model's linear relation from Figure~\ref{fig:CRE}.
Since energy is proportional to cumulative revenue even for Google, we used the historical revenue data and obtained the energy estimates. 

This highlights another use of our model, for back-casting and modeling energy use with the use of alternative data sources.
The lack of detailed and complete energy data is a general and pervasive problem.
The emissions accounting may also not be consistent across companies or over time, because of changing guidelines and accounting practices. 
However, financial metrics have a much higher availability and their relation with energy use can be used for longer-term energy models, which the thermodynamics model enables.
Thus another useful feature of our model is that it can be used with multiple data sources for checking the statistical consistency of emission data.

\section{Thermodynamic Model Implications}
\label{sec:discuss}

The use of a well-understood thermodynamics model for examining the cloud's energy usage leads to many new directions and challenges for sustainable computing research.

\subsection{Inertia and Energy Flux in Cloud Applications and Systems}
Recall the fundamental feedback loop: the efficiency causes an increase in the system size and its interface with the environment, resulting in a larger energy and material flux. 
In this subsection, we present some examples of why this is relevant and prevalent in modern cloud applications and systems.
In the cloud context, we can consider computational processing tasks and data as the equivalent of energy and matter in physical systems.
Thus, we can relate the growth of such tasks and data to the energy efficiency and consumption. 

\noindent \textbf{Demand Elasticity} is often a common reason for this growth~\cite{woodruff_when_2023}. 
Energy costs are the primary operational expense of cloud platforms, which are reduced by efficiency improvements.
A significant portion of cloud services are free of cost because of the advertisement-revenue based business model, which implies that the free cloud services have ``infinite'' elasticity as long as their use is not constrained by some other factors.
That is, the use of such free services continues to increase, both in the number of users and the modalities and quantities of use.
The increase in usage growth increases the amount of data and the associated processing tasks, which completes the feedback loop.
For example, as more users engage with the system, there are increased opportunities for analytics and bigger recommendation models. 
A second compounding feedback loop also emerges due to the pernicious nature of the advertisement-driven model: as a service attracts more usage, it leads to more data collection, which increases user-targeting capabilities, driving up revenue (i.e., thermodynamic work) and expanding the system further.

\noindent \textbf{Sub-linear Scaling} is another factor which drives system growth. 
As the system grows, more work is performed in a distributed manner. 
This leads to classic computational inefficiencies such as Amdahl's law, slowdown in distributed data processing due to communication overheads~\cite{mcsherry2015scalability}, etc. 
More recently, the growth in ML models such as LLMs necessitates highly distributed inference and newer GPUs.
Paradoxically, the sublinear scaling \emph{reduces} the efficiency, but is often a macro result of an increase in efficiency at the lower layers.

\subsection{Incorporating Clients into the System}

The energy consumption of cloud platforms drives the growth of \emph{users} and their engagement with cloud services. 
The client-side energy impact can be significant, and considering client, network, and data centers as part of the cloud system can allow similar modeling insights. 
Even though cloud platforms may use green energy and have low carbon footprint, the aggregate cloud plus users system has a high environmental impact.
The exclusive focus on making data centers efficient and ``green'' can increase the overall energy consumption and the carbon emissions, since client devices are powered by the regular electricity grid, which has a high carbon intensity globally.

\begin{figure}
  \centering
  \includegraphics[width=0.4\textwidth]{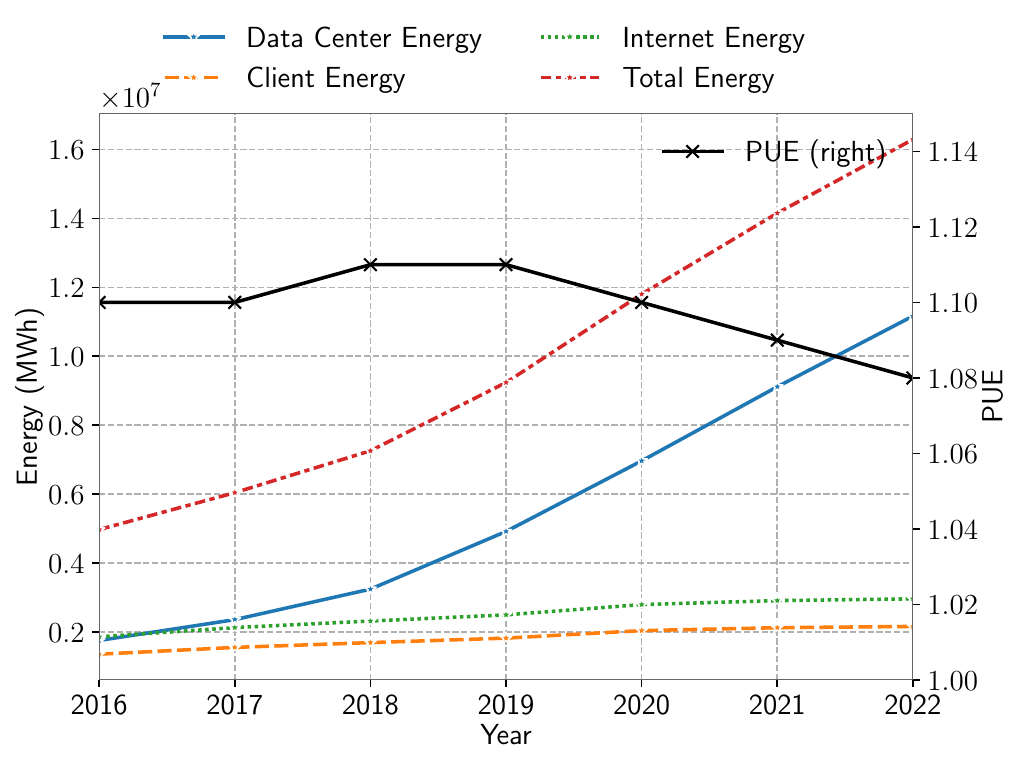}
  \vspace*{-7pt}
  \caption{The total system energy of the Meta data centers and the clients. The total client energy consumption can be significant compared to the data center energy use. For client mobile devices, we assume 2 Watt-Hours per day. Mobile chargers also have poor energy efficiency~\cite{patterson_energy_2024}.}
  \label{fig:fb-energy}
  \vspace*{-10pt}
\end{figure}

We illustrate this in Figure~\ref{fig:fb-energy} which shows the increase in the total energy consumption (in MWh) of Meta's data centers and the clients. We assume 2 hours a day of usage, with 1 Watt energy usage, which is the average usage of Facebook phone apps~\cite{phone-energy-19}. 
The internet energy is based on~\cite{inet-energy-20}, and we assume a small 10\% of internet traffic for all users being used for using these apps. 
Phone chargers are also inefficient~\cite{patterson_energy_2024}, which increases the non data center energy consumption further.
Furthermore, the embodied emissions of client devices is significantly higher than their operational emissions~\cite{gupta_act_2022}, and increasingly powerful clients are being required, exacerbating the problem further.
The carbon footprint of IoT devices~\cite{wang_low_2023} has also recently been shown to be significant, with large growth expected. 
Future work by the community can help improve the estimation and modeling of the aggregate client-side energy.

\subsection{Characterizing Efficiency and Work}

Metrics often provide the incentives for system evolution and improvements.
For today's hyperscale clouds, our empirical analysis indicates that revenue may be the canonical metric for work.
However, for finer-grained modeling and characterization, we will need workloads and metrics which capture the end-to-end cloud use.
Since the clouds are complex adaptive systems, we need to measure and define efficiency across multiple scales.

This will require the community to construct and update portfolios of common operations and applications with their total client-side, networking, and cloud energy impact.
This portfolio can include common tasks like sending an email, opening and editing a document, etc.
Tasks of the right granularity will be key: hardware efficiency improvements (Moore's law etc.) can provide some benefits, but software abstractions and bloat can offset these.
For example, the energy required for running an IMAP server and client on old hardware, vs. an entire Exchange infrastructure with heavy clients and browser javascript, may yield interesting tradeoffs.
Since the energy usage and carbon footprint of clients can be significant (illustrated in Figure~\ref{fig:fb-energy}), metrics for end-to-end efficiency will be key to sustainable cloud computing. 

\noindent \textbf{Role of efficiency: What to optimize?}
The deeper understanding of the mechanisms of Jevons paradox, such as the role of inertia and network effects, can also help guide optimization efforts which minimize rebound effects. 
The model implies that optimizations at the ``edge'' of the system will have most impact, and external constraints are crucial to prevent run-away growth.
Thus client-focused efficiency optimizations can improve both the user experience (improve UI latency) and sustainability, and should be compared against conventional PUE-improving have data center optimizations.

\subsection{Alternate Cloud Architectures}

\noindent \textbf{Distributed edge clouds} may impose location constraints and natural boundaries, which limit system growth.
From the thermodynamics model perspective, if each distributed edge computing node serves only a limited set of users, then it can be examined as a self-contained system.  
This may provide different sustainability tradeoffs---their operational efficiency metrics such as PUE may be worse than hyperscale data centers, but the overall energy use may be lower. 
Longer-term empirical analysis of edge computing will be necessary to understand these tradeoffs in distributed computing architectures.

\bibliographystyle{acm} 
\bibliography{jevons,carbon}
\end{document}